\documentclass{article}
\usepackage[dvips]{graphicx}

\usepackage{latexsym}

\begin{document}

\pagestyle{plain} 
\setcounter{page}{1}
\setlength{\textheight}{700pt}
\setlength{\topmargin}{-40pt}
\setlength{\headheight}{0pt}
\setlength{\marginparwidth}{-10pt}
\setlength{\textwidth}{20cm}

\title{Second  Parrondo's Paradox in Scale Free Networks}
\author{Norihito Toyota   \and Hokkaido Information University, Ebetsu, Nisinopporo 59-2, Japan \and email :toyota@do-johodai.ac.jp }
\date{}
\maketitle

\begin{abstract}
Parrondo's paradox occurs in sequences of games in which a winning expectation value of a payoff 
may be obtained by playing two games in a random order, 
even though each game in the sequence may be lost when played individually. 
Several variations of Parrondo's games apparently with the same paradoxical property have been introduced \cite{Harm3};
 history dependence, one dimensional line,  two dimensional lattice and so on. 
I have shown that Parrondo's paradox does not occur in scale free networks in the simplest case 
with the same number of parameters as the original Parrondo's paradox\cite{ToyotaPa}. 
 It suggests that some technical complexities are needed to present Parrondo's paradox in scale free networks. 
 In this article, I show that a simple modification with the same number of parameters as the original Parrondo's paradox 
creates Parrondo's paradox in scale free.  
This paradox is, however, created by  a quite different mechanism from the original  Parrondo's paradox and a considerably rare phenomenon,   
where the discrete property of degree of nodes is crucial. 
I call it the second Parrondo's paradox. 
 \end{abstract}
\begin{flushleft}
\textbf{keywords:}
 Parrondo's paradox, Parrondo's paradox，Scale  free network, Game theory
\end{flushleft}

\section{Introduction}\label{intro}
\hspace{5mm} 
Parrondo's games were first devised by Parrondo \cite{Parr1} who presented them in unpublished.   
Parrondo's paradox is based on the combination of two losing games. 
However, the randomly combined two games give rise to a winning game pointed out by \cite{Harm1} and \cite{Harm2}. 
It was  pointed out that Fokker Planck equation connects  Parrondo's games with the flashing Brownian ratchet \cite{Amen}. 
The original Parrondo's game consists of two losing games A and B where each is played by only one player.  
In Game A, only one biased coin is used, while in Game B two biased coins are used with the player's current capital determining the state 
dependent rule. 
When a player plays individually each game, he/she loses his/her capital on average. 
However, when the player plays two games in any combination, he/she always wins on average.    

Several variations of Parrondo's games apparently with paradoxical property have been introduced. 
First capital dependence in Game B was replaced by recent history of wins and loses \cite{Parr2}. 
It has the same paradoxical property as the original one. 
For it, an analytical study has been made \cite{Rasm1}. 
Secondly the capital dependence in Game B was replaced by spatial neighbor dependence 
in one dimensional line \cite{Toral1}, \cite{Miha1}, \cite{Miha2},  and von Neumann  neighbor dependence in two dimensional lattice \cite{Miha}.  
Some analogous variations have been given by  \cite{Toral2},\cite{Davi}. 
Among such positive studies, I have especially shown that Parrondo's paradox does not occur in scale free networks in the simplest case 
with the same number of parameters as the original Parrondo's paradox\cite{ToyotaPa}. 
This fact suggests that some technical complexities are needed to present Parrondo's paradox in scale free. 
A fine review of Parrondo's paradox and references except for my work are given in \cite{Harm3}. 

In this article, I explore still harder about Parrondo's paradox in scale free networks\cite{Albe1},\cite{Albe2}. 
The study of  Parrondo's paradox in scale free networks is interesting as an empirical study, 
since scale free networks are ubiquitous in our real world\cite{newBook}.  
 In this article I show that a simple modification to the games with the same number of parameters 
as the original Parrondo's paradox creates Parrondo's paradox in scale free networks. 
In the model of this article, each player on a network plays a game L when there are not more than $rk_i$ winners in the neighborhood 
connected to the player and plays a game W otherwise in Game B, 
where $k_i$ is the degree of node $i$ and $r$ is a parameter taking between $0$ and $1$. 
As the theoretical studies, first the conditions for Parrondo's paradox are obtained.   
I give some simulation result which remarkably shows a paradox occurs based on the conditions.  
But another theoretical consideration shows that such paradox never occurs.  
I examine carefully those considerations.  
As a result, I find that Parrondo's paradox can occur in the present naive case on scale free networks.   
This paradox is, however, created by  a quite different mechanism from the original  Parrondo's paradox. 
The occurrence of the paradox is very subtle and a considerably rare phenomenon, where the discreteness of degree of nodes is crucial. 
I call it the second  Parrondo's paradox because it is created by  a quite different mechanism. 

\section{Review of Parrondo's Paradox}
\subsection{Original Game}
\hspace{5mm} 
I give a brief review of the original Parrondo's game in this section. 
Parrondo's paradox occurs in sequences of games in which a winning expectation may be obtained by playing two games in a random order, 
even though each game in the sequence may be lost when played individually.  
The original version of Parrondo's game consists of  the following two games; \\
\begin{itemize}
\item Game A: the probability  of winning is $P_A$ in this game. $P_A<0.5$ is taken for losing game. 
\item Game B: If the capital $C(t)$ of the player at $t$ is a multiple of 3, the probability of winning is $P_B^{(1)}$, otherwise, the probability of winning is $P_B^{(2)}$
\item Game A+B: Two games are mixed. The game A is played with probability $P$ and Game B is played with $1-P$.
\end{itemize}
In all there are 4 parameters,  $P$, $P_A$, $P_B^{(1)}$ and $P_B^{(2)}$,  controllable by  a planner of the game  in Parrondo's game.  
When we win Game A or B or A+B,  we get one unit of capital and when we lose the game, we lose one unit of capital.

Time series of $C(t)$ of Game A and Game B are given by Fig.1, respectively, where $C(0)=100$ is taken. 
They show  that  the capital $C(t)$ decreases with time $t$ in both the games. 
Fig 2, which displays the time series of the game A+B, however, shows that   the capital $C(t)$ increases with time $t$.  
This is an example of Parrondo's paradox. 

 \begin{figure}[t]
\begin{center}
\includegraphics[scale=0.7,clip]{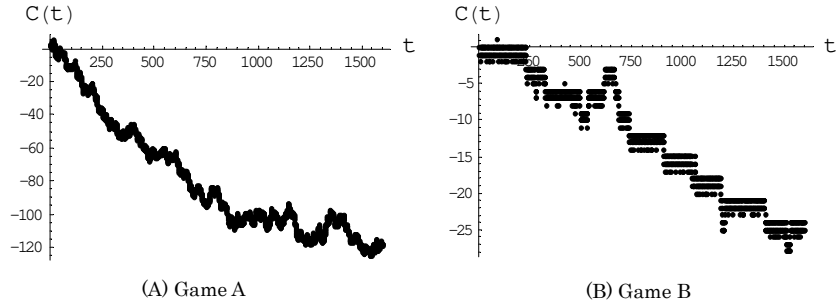} 
\end{center}
\caption{$C(t)$ in Game A and Game B for $P_A=0.48$,  $P_B^{(2)}=0.01$ and $P_B^{(1)}=0.85$ at 1600 steps.  }
\vspace{4mm}
\end{figure}

 \begin{figure}[t]
\begin{center}
\includegraphics[scale=0.7,clip]{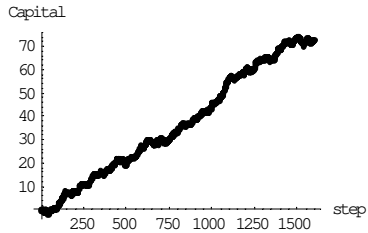} 
\end{center}
\caption{$C(t)$ in Game A+B for $P=0.5$, $P_A=0.48$,  $P_B^{(2)}=0.01$ and $P_B^{(1)}=0.85$ at 1600 steps.}
\end{figure} 
 
 This paradox can be understood by an analysis using a difference equation for the probability 
that the capital reaches zero in a finite number of plays when we initially have a given capital of $j$ units\cite{Amen},\cite{Harm3}.  

The game A is classified into the following three cases as  result;
\begin{eqnarray}
D_{(A)} ^{(3)} < 1  &so& P_A>\frac{1}{2}\;\;\;\;\;\mbox{ winning game},\\
D_{(A)}^{(3)} = 1    &so& P_A=\frac{1}{2} \;\;\;\;\;\mbox{ fair game},\\
D_{(A)}^{(3)}   > 1  &so &  P_A<\frac{1}{2}    \;\;\;\;\;\mbox{ losing game},
\end{eqnarray} 
where 
\begin{equation}
D_{(A) }\equiv  \frac{1-P_A}{P_A}.   
\end{equation}

Game B is also classified into the following three cases by the similar way;
\begin{eqnarray}
D_{(B) } ^{(3)} < 1  &\;\;\;\;\;\mbox{ winning game},\\
D_{(B) } ^{(3)} = 1    &\;\;\;\;\mbox{ fair game},\\
D_{(B) } ^{(3)}  > 1  &\;\;\;\;\mbox{ losing game},
\end{eqnarray} 
where
\begin{equation}
D_{(B) }  \equiv   \frac{ (1-P_B^{(1)})(1-P_B^{(2)})^2}{ P_A(P_B^{(2)})^2 }.    
\end{equation} 

 Game A+B where Game A is chosen randomly with the probability $P$ or Game B  with the probability $1-P$. 
I classify the game into two cases  as in Game B.   
   \begin{itemize}
\item The case that the capital is a multiple of three. 
        The winning probability is 
\begin{equation}
 P_{A+B}^{(1)} \equiv  PP_A + (1-P)P_B^{(1)}.   
\end{equation}  
\item The case that the capital is not a multiple of three 
        The winning probability is 
\begin{equation}
 P_{A+B}^{(2)} \equiv  PP_A + (1-P)P_B^{(2)}.   
\end{equation}  
\end{itemize}
So it is found that Game A+B is included in Game B as a special case by the following replacement;  
\begin{equation}
 P_{B}^{(1)} \Longrightarrow  P_{A+B}^{(1)} \;\;\;\;\;\;P_{B}^{(2)} \Longrightarrow  P_{A+B}^{(2)}.   
\end{equation}  
When I define the following $ D_{(A+B)}^{(3)} $,  
\begin{equation}
  D_{(A+B)}^{(3)} \equiv \frac{ (1-P_{A+B}^{(1)})(1-P_{A+B}^{(2)})^2}{ P_{A+B}^{(1)}(P_{A+B}^{(2)})^2 },  
\end{equation}
Game A+B is classified  such as Game B 
\begin{eqnarray}
 D_{(A+B)}^{(3)}  < 1  &\;\;\mbox{ winning game},\\
 D_{(A+B)}^{(3)} = 1    & \;\;\mbox{ fair game},\\
 D_{(A+B)}^{(3)} > 1  &\;\;\mbox{ losing game}.
\end{eqnarray} 

The game A can be also considered as Game B with $P_B^{(1)}=P_B^{(2)}=P_A$. 
Considering it together with eq. (11), both Game A and Game A+B are essentially special cases of Game B.  
Thus I can represent the discriminants in the unified way by replacing the subscript B with A or A+B  in eq. (5)-(8).

When the conditions 
\begin{eqnarray}
 D_{(A)}^{(3)}  > 1, &
 D_{(B)}^{(3)} >  1, & 
 D_{(A+B)}^{(3)} < 1 
\end{eqnarray} 
are simultaneously satisfied, Parrondo's paradox occurs. 

\subsection{Some Extensions of Parrondo's Paradox }
\hspace{5mm} 
Let consider the a more general case that  when a capital satisfies  $mod(C(t),M)= 0$ for a positive integer $M$,  
a game with a winning probability $P_B^{(1)}$ is played and the another  game with a winning probability $P_B^{(2)}$ is played 
 in the cases of  $mod(C(t),M) \neq 0$ in Game B. 
Then the  discriminant is given by 
\begin{equation}
  D_{A+B}^{(n)}= \frac{ (1-P_{A+B}^{(1)})(1-P_{A+B}^{(2)})^{M-1}}{ P_{A+B}^{(1)}(P_{A+B}^{(2)})^{M-1} }.     
\end{equation}
Then it is known that the paradox occurs under the conditions like eq. (16)\cite{Harm3}. 

Such paradox also occurs in some extended versions of the original Parrondo's game. 
One of them is that two games  in Game B are chosen depending on historical information as to winning-losing  
\cite{Parr2}. 
Some theoretical discussion are given for the game\cite{Rasm1}. 

The case that players lying on one dimensional circle lattice play Parrondo's game as an other case\cite{Toral1},\cite{Miha1},\cite{Miha2} is investigated. 
Then two games in Game B are chosen depending on information as to winning-losing of two adjacent players ( left and right of the target player). 
Moreover the case is extended to two dimensional regular lattice with the periodic boundary condition\cite{Miha}.  
It has been verified that the paradox can also occur in these cases\cite{Toral1},\cite{Miha}. 

Next I briefly review the extended case to two dimensional lattice according to Mihailovic et al. \cite{Miha}. 
They introduce five subgames in Game B, depending on one winner, two winners, three winners, four winners in von Neumann neighborhood of a target player. 
The target player plays one of games with the winning probability of $p_B^{(0)}$，$p_B^{(1)}$，$p_B^{(2)}$，$p_B^{(3)}$ and $p_B^{(4)}$ corresponding to 
 the number of winners of his/her adjacent player. 
So Game B has five parameters. 
They analyze  the average capital over all players as the time series of capitals.  
As result, they report that the paradoxical phenomena occur in wide range of parameter sets, 
especially it appears typically in synchronous cases where all players play at the same time.   

\section{Parrondo's Paradox on Scale Free Networks I}

\subsection{Parrondo's Game I on Scale Free Networks (Threshold Game)}
\hspace{5mm} 
The degree distribution of scale free networks with the lowest degree $k_{min}$ is given by 
\begin{equation}
P(k)= ck^{-\alpha} = \frac{k^{-\alpha}}{\zeta (\alpha, k_{min})}.
\end{equation}
where $\alpha>0$ is an exponent in a power low, $c$ is a normalization constant and the Hurwitz zeta function
is defined by
\begin{equation}
\zeta (\alpha, k_{min}) \equiv \sum_{n=0}^\infty (n+k_{min})^{-\alpha}.  
\end{equation}

In this article, I construct scale free networks by using the preferential attachment (BA model) introduced by Barabashi et al. \cite{Albe1},\cite{Albe2} 
where the scaling exponent is theoretically $\alpha=3$
(Barabasi et al. have actually shown that $\alpha=2.9\pm 0.1$ for BA model with a few$\times 10^5$ nodes\cite{Albe1},\cite{Albe2} ).      
In this article, I begin with the complete graph with degree 4,  nodes with degree 4 to the initial graph are attached by 
using BA model up to the network size $N$. 

I consider a naive extension of Parrondo's game on scale free networks. 
Parrondo's game is straightforwardly extended to the game with degree $k$, which means the player on the node 
with degree $k$ plays against $k$ players, according to \cite{Miha} such as explained in 2.2. 
Then the number of parameters appearing as the winning probabilities  becomes $k+1$ and that is so large. 
The reason is that there can be $0 \sim k$ winers in the adjacent to the target node with degree $k$. 
When the target player plays different games depending on the number of winners adjacent to him/her, 
the discriminant of the node with degree $k$ is generally given by 
\begin{equation}
 D_{(k,p_i)}= \frac{ \displaystyle\prod_{i=0}^{k} (1-p_{i,B}) }{ \displaystyle\prod_{i=0}^{k} p_{i,B} },   
\end{equation}
where $p_{i,B}$ is the winning probability in the case that there are $i$ winners adjacent to the target player 
 in GameＢ. 
So $k+1$ games are necessary. 
Furthermore every node has different degree in scale free networks, and so by all accounts, such game has too many parameters 
to analyze the game theoretically. 

So I introduce a cutoff $R$ as a simple idea to analyze the game on scale free networks.   
When there are not more than $R$ winners adjacent to a target player, the target player plays Game L whose winning probability is $P_{L,B}$, 
 and the target player plays Game W whose winning probability is $P_{W,B} $ in other cases  in Game B. 
Then the  number of parameters of the games is the same as the original Parrondo's game due to this simplification.  
So the discriminant for the node with degree $k$ for Game B in this Parrondo's game on scale free networks 
based on analogical inference of eq.(17) is 
\begin{equation}
D_{(k,P_{W,B} ,P_{L,B}  )}= \frac{ (1-P_{W,B})^{k-R} (1-P_{L,B} )^R }{ P_{W,B}^{k-R} P_{L,B}^{R} }.    
\end{equation}
 
We need only to  replace the subscript $B$ with $A+B$  for Game A+B where  
the relations of winning probability between both games are essentially given by eq.(9) and eq.(10).  

Since Parrondo-like paradox may accidentally occur in the capital of individual because of probability game,  
I focus my analyses on the average capital over all players as the time series of capitals 
in similar manner to the game on two dimensional lattice\cite{Miha2}.

\subsection{Theoretical and Numerical Studies }
\hspace{5mm}
In this case, it may be effective that the theoretical analysis is made 
via DTMC(discrete-time Markov chains) such as given in \cite{Harm3},\cite{Miha1},\cite{Miha2}. 
As exploring Parrondo's paradox following this analysis, there is a difficulty,  because every node, however, 
has a different degree in a scale free network and so the size of the transfer matrix can not be fixed. 
 Therefore the analysis based on DTMC is difficult for scale free networks.   
 I make the analysis in this case based on the mean field approximation such as a recursion relation. 
 
 Let's consider the condition that some paradox occurs in this game.   
Notice that when the degree of  nodes  is smaller than the cutoff $R$, the players on the nodes necessarily play 
Game L in Game B. 
Then the discriminant is given by 
\begin{equation}
D_{(k,P_{L,B} )}= \frac{ (1-P_{L,B} )^k }{ P_{L,B}^{k} }.   
\end{equation}

 \begin{figure}[t]
\caption{ ($P_{W,B}$，$P_{L,B}$) that satisfies the condition eq.(23) and (24) at $R=4$ and $R=9$ for $k_{max}=25$, 
at $R=30$ for $k_{max}=65$ and at $R=73$ for $k_{max}=150$.     }
\end{figure} 

 \begin{figure}[t]
\begin{center}
\includegraphics[scale=0.9,clip]{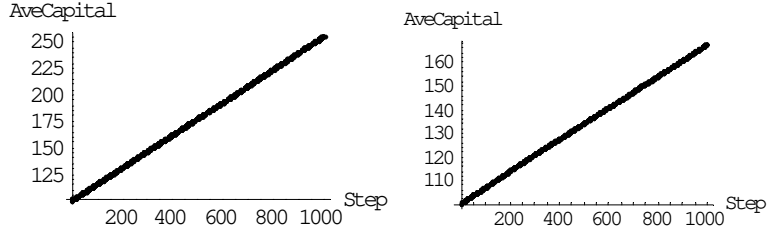} 
\end{center}
\caption{ Average capital $C(t)$ and capital distribution in every node in Game B (left) and Game A+B (right)  
at $R=10$ and $(P_{L,B}, P_R)=(0.6,0.2)$.  }
\end{figure}

Players on nodes with high degree can play either Game W or Game L corresponding to the number of winners in neighboring players.
Taking account of $k_{min} =4$ in BA model adopted in this article, 
Using the expectation values of the discriminant, I find the following conditions for the occurrence of the paradox in the continuous approximation;  
\begin{eqnarray}
\langle D_{(k,P_{W,B} ,P_{L,B}  )} \rangle =\Bigl( \int_{4}^R  \; \frac{D_{(K,P_{L,B} )}}{ K^{\alpha} }dK +  
\int_{R+1}^k   \frac{D_{(K,P_{W,B} ,P_{L,B}  )}}{ K^{\alpha} }dK \Bigl) 
 &\times    \frac{(\alpha -1)}{ k_{min}^{1-\alpha} - k_{max}^{1-\alpha} }<1, \\
\langle D_{(k,P_{A+B} )} \rangle =  \Bigl(  \int_{4}^R  \; \frac{D_{(K,P_{L,A+B} )}}{  K^{\alpha}  }dK +
 \int_{R+1}^k   \frac{D_{(K,P_{W,A+B} ,P_{L,A+B}  )}}{ K^{\alpha} } dK\Bigl) 
 \times &   \frac{(\alpha -1)}{ k_{min}^{1-\alpha} - k_{max}^{1-\alpha} } >1,
\end{eqnarray}
where $c=\frac{(\alpha -1)}{ k_{min}^{1-\alpha} - k_{max}^{1-\alpha} } $ in the  continuous approximation\cite{ToyotaPa}. 
I will investigate these conditions by making numerical calculations. 

Fig.3 shows the typical data points that satisfy the conditions eq.(23) and eq. (24) at various $R$, $P=0.5$ and $N=400$.   
I confirmed that the data points with large $P_{W,B} $ in the left end of Fig.3, which appear at small $R$, 
disappear as $k_{max}$ becomes larger. 
As for large $R$, data points appear in some straight lines shown in  Fig.3. 
These points appear at still larger $R$ (so necessarily large $k_{max}$) as shown in the right end of Fig.3.  
As $R$ becomes larger, the gradient becomes smaller and the number the total data points reduces.  
From a series of these figures,  it is conjectured that the data points in the straight lines will disappear 
at the limit of $N \rightarrow \infty$, which would be  due to a sort of the finite size effect. 
From these facts, I surely speculate that Parrondo's paradox does not occur on scale free networks 
with the cutoff game which has the same number of parameters as the original one.  

Moreover I  put to the test by a computer simulation at $N=400$ and $P=0.5$  in order to ensure the conjecture.
The region that should be explored in the parameter space $P_{L,B}$-$P_{W,B}$ for the paradox are limited to   
\begin{equation}
(P_{L,B}-0.5)(P_{W,B}-0.5)<0. 
\end{equation}
This is because of $P_{L,B}  \Leftrightarrow P_{W,B}$ duality symmetry in the present case. 
Computer simulations are made  under this parameter region where each player on the network asynchronously plays with $C(0)=100$.    
 In the case of $R=10$ and $P_{L,B}>0.5>P_{W,B}$, a typical time series of an average capital over all players at $t=1000$ 
 are shown in Fig.4 for Game B and  Game A+B. 
In general, the average capital $C_B(t)$ in Game B is lager than  $C_{A+B}(t)$ in Game A+B. 
So it would be difficult that the average capital in Game A+B  exceeded the one in Game B. 

 In the case of $R=10$ and $P_{L,B}<0.5<P_{W,B}$, there are not any essential differences from the above cases 
except that both the  $C_{B}(t)$ and  $C_{A+B}(t)$ decrease due to the duality of $P_{L,B}  \Leftrightarrow  P_{W,B}$.  
Furthermore, the variations of $P_{L,B}$ and $P_{W,B}$ within above two parameter regions have an effect on only the variation 
of the absolute values of   $C_{B}(t)$ and $C_{A+B}(t)$ but does not occasion any qualitative changes such as a reversal in capital. 
I can actually show  that decreasing $R$ makes losers increase in the case  of $R=10$ and $P_{L,B}>0.5>P_{W,B}$, 
but makes losers decrease in the case  $R=10$ and $P_{L,B}<0.5<P_{W,B}$\cite{ToyotaPa}. 
These circumstantial evidence shows that the variation of $R$ contributes only to the absolute value of the capital, 
and does not bring about any qualitative change such as the occurrence of paradox.  

As result of all simulations, I also found that paradox does not occur in these data points at these $R$. 
It is a future problem whether paradox actually occurs at exceedingly larger $R$ in large scale networks. 
But taking account of also theoretical considerations, 
I surely speculate that Parrondo's paradox does not occur in the threshold game on scale free networks.

\section{Parrondo's Paradox on Scale Free Networks II}

\subsection{Parrondo's Game II on scale free networks (threshold ratio game)}
\hspace{5mm} 
I consider another naive extension of Parrondo's game on scale free networks. 
I prepare a cutoff for winners' ratio $r$ rather than the number of winners adjacent to a target player.    
When there are not more than $rk_i$ winners adjacent to a target player $i$, the target player plays Game L 
whose winning probability is $P_{L,B}$, 
 \begin{figure}[t]
\begin{center}
\includegraphics[scale=0.9,clip]{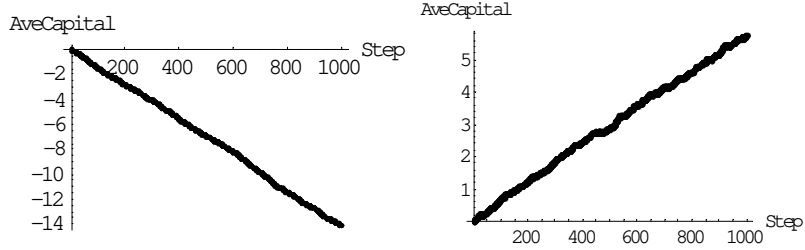} 
\end{center}
\caption{The average capital $C(t)$ ( $C(0)=100$ ) at  $(P_{L,B}, P_{W,B},r)=(0.68,0.092,2/3)$ in Game B (left) and Game A+B (right) in the threshold ratio game where the network size is 1000.  }
\end{figure} 
 and the target player plays Game W whose winning probability is $P_{W,B} $ in other cases  in Game B. 
The  number of parameters of the games is also the same as the original Parrondo's game due to this simplification.
Then the discriminants for the node with degree $k$ for Game B and Game A+B are
\begin{eqnarray}
D_{(k,P_{W,B} ,P_{L,B}  )}&=& \frac{ (1-P_{W,B})^{k-R} (1-P_{L,B} )^R }{ P_{W,B}^{k-R} P_{L,B}^{R} }, \\
D_{(k,P_{W,A+B} ,P_{L,A+B}  )}&=& \frac{ (1-P_{W,A+B})^{k-R} (1-P_{L,A+B} )^R }{ P_{W,A+B}^{k-R} P_{L,A+B}^{R} },
\end{eqnarray}
where $R=rk_i$.  
The conditions that some paradox occurs are
\begin{eqnarray}
D_{(k,P_{W,B} ,P_{L,B}  )}&>&1,  \\
D_{(k,P_{W,A+B} ,P_{L,A+B}  )}&<& 1,
\end{eqnarray}
where $P_A<0.5$ and $P=0.5$ is taken. It is natural because Game A is losing.  
When  the ratio of  winners  to total players adjacent to target one is $r$, (28) and (29) become
\begin{eqnarray}
 \frac{1}{ 1+(\frac{P_{W,B} } {1-P_{W,B} } )^\beta }    &>& P_{L,B},\;\;\;\;\;\mbox{ for }P_{L,B} <\frac{1}{2} < P_{W,B},\\
  \frac{1}{ 1+(\frac{P_{L,B} } {1-P_{L,B} } )^{\frac{1}{\beta} } }    &>& P_{W,B},\;\;\;\;\;\mbox{ for }P_{W,B} <\frac{1}{2} < P_{L,B},
\end{eqnarray}
 and
\begin{eqnarray}
 \frac{1}{ 1+(\frac{P_{W,A+B} } {1-P_{W,A+B} } )^\beta }    &<& P_{L,A+B},\;\;\;\;\;\mbox{ for }P_{L,A+B} <\frac{1}{2} < P_{W,A+B},\\
  \frac{1}{ 1+(\frac{P_{L,A+B } }{1-P_{L,A+B} } )^{\frac{1}{\beta} }  }    &<& P_{W,A+B},\;\;\;\;\;\mbox{ for }P_{W,A+B} <\frac{1}{2} < P_{L,A+B},
\end{eqnarray}
respectively,  where $\beta$ defined by the below does not depend degree $k$;
 \begin{equation}
 \frac{k-R}{R}=\frac{k-rk}{rk}=\frac{1-r}{r} \equiv \beta.
 \end{equation}
 
 First of all, I discuss  (31) and (33). The necessary conditions for satisfying both inequations are
\begin{eqnarray}
\frac{1}{ 1+(\frac{P_{L,B} } {1-P_{L,B} } )^\frac{1}{\beta} }    &>& P_{W,B},\\
 P_A+P_{W,B}  &>&1, \\
 P_{L,B}  >\frac{1}{2} &>&P_{W,B} .
\end{eqnarray}
The below parameter set satisfies the condition  (35), (36) and (37);
\begin{equation}
(P_A, P_{W,B},P_{W,L},r,P)= (0.49,0.1,0.68,2/3,0.50).
\end{equation}  
So they are candidate parameters for a paradox. 
Fig. 5 really shows that a paradox in the average capital occurs at the parameter set 
where $C(0)=0$ where the network size $N=1000$ are taken.

However, since the conditions for the Parrondo's paradox have not any $k$ dependence from (34), 
As result, the wining probabilities for Game B and Game A+B are given by  
\begin{eqnarray}
P_B^{eff} &=&s P_{W,B} +(1-s) P_{L,B}, \\
 P_{A+B}^{eff} & =& pP_A+ (1-p) \bigr( sP_{W,B} +(1-s)P_{L,B} \bigl), 
\end{eqnarray}
where $s$ is the possibility game W is chosen.
Initially game L or game W is randomly and uniformly assigned for each player on nodes. 
The possibility that the adjacent players to a player on a node play game L or game W is given by 
a binomial distribution.  
When a cut off $r$ is introduced, $s$ is evaluated through a cumulative value of the binomial distribution. 
 
Remember that two kinds of probabilities for the occurrence of the paradox are needed for any nodes 
such as in the original Parrondo game B. 
There is, however, one control parameter $s$ in (39)  due to $k$ independence.  
Thus it seems that there is  no sign of the paradox in this case. 
In fact substituting (39) into (40), we obtain
\begin{equation}
 P_{A+B}^{eff} =p P_A+ (1-p) P_{B} = p(P_A-P_B)+P_B. 
\end{equation}
So we find $ P_{A+B}^{eff} <0.5$ for $p=0.5$ or even $p<1$, $P_A<0.5$ and $0.5>P_B$. 
These considerations clearly show that both game B and game A+B are losing.    

I investigate why the Parrondo's paradox such as Fig.5 occurred.   
Notice that effective $r$ is always some rational number different from true $r$ for a special $k$, 
since $k$ can take only a natural number.
So $r$ takes some effective value $r_{eff}$ due to the discrete nature of $k$.  
For example, the possible winning ratios of a node with $k=4$ are $0/4$, $1/4$, $2/4$, $3/4$ and $4/4$. 
So choosing $r=2/3$,  $r$ becomes effectively $r_{eff}=r_{upper}=3/4$ for W game.    
 For L game, $r$ likewise becomes effectively $r_{eff}=r_{down}=2/4$. 
Thus two effective $r$'s appear. 
From the fact, evaluating  $s$ based on a cumulative binomial distribution, we find $s\sim 0.6875$. 
We obtain $s\sim 0.8125$ for $k=5$ in a similar way.    
Thus $s$ depends on a degree $k$, so that $s$ changes every $k$ in (39). 
Consequently the discreteness of $k$ induces several $P_B^{eff} $'s  which trigger paradoxical phenomena.     

For the same parameters as ones in Fig.5, the effective values of  $P_B^{eff} $ and $P_{A+B}^{eff} $ are shown in Fig.6. 
 \begin{figure}[t]
\begin{center}
\includegraphics[scale=0.9,clip]{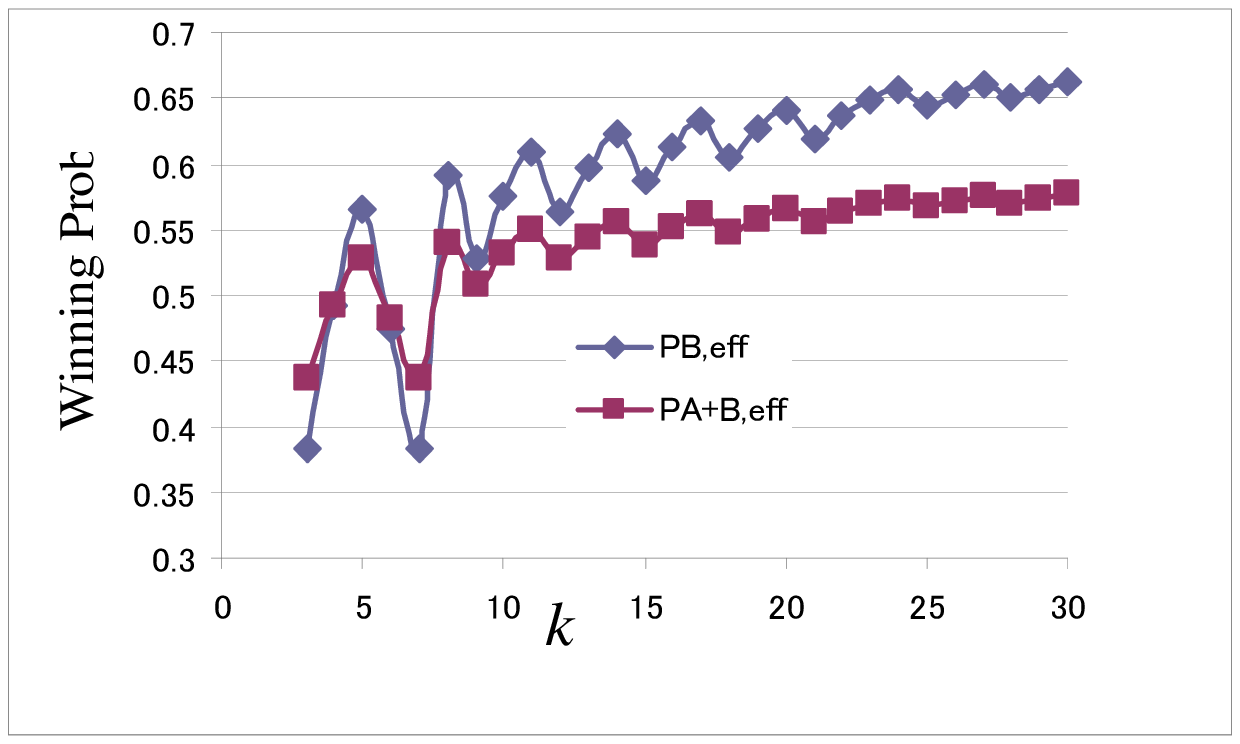} 
\end{center}
\caption{$P_B^{eff} $ and $P_{A+B}^{eff} $ for various $k$ values in the parameter (38) .  }
\end{figure} 
In Fig. (6),   $P_B^{eff} $ is sporadically smaller than  $P_{A+B}^{eff} $ at small $k$. 
This region of $k$ gravely affects the average values $<P_{B,eff}>$ and $<P_{A+B,eff}>$  because of the nature of scale free networks. 
The average values $<P_{B,eff}>$ and $<P_{A+B,eff}>$ over $k$ are given by
\begin{equation}
 <P_{B(A+B)}^{eff} >= \int P(k) P_{B(A+B)} ^{eff} (k)=\int ck^{-\alpha} P_{B(A+B)}^{eff} (k).
 \end{equation}
Since this value is sensitive to the normalization constant $c$, we refine parameters in the present model.   
I first use $\alpha =2.57 $ (driven based on the least squares method) 
for the network really constructed in this article, different from ideal value $\alpha=3$. 
Though $k$ can ideally take infinite large value in itself, $k$ takes from $k_{min}=3$ 
  to $k_{max}=30$\cite{ToyotaPa}  in Fig.6.  
 
Then the theoretical value of the normalization constant $c$ of the degree distribution can be evaluated. 
 But when the normalization constant is corrected to $1.053$ times of 
the observed value in the network with $k_{max}=30$, we obtain $<P_{B,eff}>\sim 0.49$ and $<P_{A+B,eff}>\sim 0.503$. 
The corresponding values of the capitals for game B and game A+B are $C_B(t=1000)\sim 81$ and 
$C_{A+B}(t=1000)\sim 106$, respectively. 
 These values almost explain the capitals of game B and game A+B in Fig. 5 well. 
 The exact theoretical correction for the normalization constant without any limit of $k$ is  
$1.035 \times$ the theoretical value at $k_{max}=30$,  which is comparable with the above value. 
   
There are effectively multiple $r_{eff}$ and $s$ , and so multiple $<P_{B,eff}>$ and $<P_{A+B,eff}>$,  
due to the discrete nature of $k$.  
Such the effect  creates Parrondo's paradox in the similar way as the original Parrondo's game,  
but not in different mechanism from the original one.   
Thus I call its paradox the second Parrondo's paradox.    

\section{Summary}
\hspace{5mm}
In this article, I explored whether paradox occurs or not in Parrondo's game on scale free networks which are more ubiquitous in real worlds than regular networks.   
It is too complicate to analyze the game in the general fashion, especially in giving theoretical considerations.  
So I consider only the case with the same number of parameters as the original Parrondo's game based on modulo $M=3$ in the capital.   
In this article, the parameter corresponding to $M$ in the original Parrondo's game  is the cutoff $R$.    
When the number of winners adjacent to a target player is not more than $R$,  
the player plays Game L with the winning probability $P_{L,B}$ in Game B of Parrondo's game. 
Otherwise the player plays Game W with the winning probability $P_{W,B}$.
Two types of models are considered in this article. 
One is the the threshold game where $R$ takes a constant value independent of degree $k$. 
Another is the threshold ratio game where $R$ depends on degree $k_i$ of node $i$ such as $R_i=rk_i$. 
$r$ means the ratio of winners to all players adjacent to the player on node $i$.   

For the threshold  game, I accumulated circumstantial evidence that paradox does not occur by some computer simulations.
Furthermore I almost practically showed  that Parrondo's paradox did not occur in this naive case 
from theoretical point with numerical experiments of view.  
It, however, remains to be studied whether some paradox actually does not occur in excessively large scale networks. 

I showed that a paradox occurred for some parameter set in the threshold ratio model. 
In this case, the discrete nature of $k$ effectively induces plural winning probabilities in game B as the original Parrondo's game. 
It is thought that the various $P_{B,eff}$ or $P_{A+B,eff}$ appearing in this model create the paradox. 
The plural winning probabilities in game B are not given artificially as  the setting of the game but they 
are effectively induced by the discrete nature of $k$. 
So I called the paradox the second Parrondo's paradox. 

But the inhomogeneity in degree is not reflected in Model II, explicitly. 
Model II looks like Parrondo's games on a lattice where Parrondo's paradox occurs\cite{Toral1}, \cite{Miha1}, \cite{Miha2},\cite{Miha}  in this point. 
Synthetically, it is considered that it is difficult to create some paradoxical behavior in models based on networks with inhomogeneous degree.

I never showed that it was hard to create Parrondo's paradox in scale free networks, generally. 
Notice that  the networks that  I studied is only BA models and only studied excessively naive setting in game B.  
I only focus my attention  on the numbers or the ratio of winners, and not on degree or the number of losers
(notice that considering both the number of losers and winners turns  out to consider degree and the number of winners as well). 
If making efficient use of  these information, we would find more rich aspects of Parrondo's  paradox in various types of scale free networks.


\small

\end{document}